\documentclass[reprint,aps,prb,twocolumn,amsmath,amssymb,showpacs,preprintnumbers]{revtex4-2}

\usepackage{graphicx}[]% Include figure files
\usepackage{dcolumn}% Align table columns on decimal point
\usepackage{bm}% bold math
\usepackage{multirow}

\usepackage{braket}
\usepackage{color}
\usepackage{ulem}
\usepackage{url}
\usepackage{hyperref}

\definecolor{myblue}{RGB}{50,50,200}

\begin{document}

\title{
Magnetic bubble crystal in tetragonal magnets
}
\author{Satoru Hayami$^1$ and Yasuyuki Kato$^2$}
\affiliation{
$^1$Graduate School of Science, Hokkaido University, Sapporo 060-0810, Japan \\
$^2$Department of Applied Physics, the University of Tokyo, Tokyo 113-8656, Japan 
}

\begin{abstract}
A magnetic bubble crystal is a two-dimensional soliton lattice consisting of multiple spin density waves similar to a magnetic skyrmion crystal. 
Nevertheless, the emergence of the bubble crystal with a collinear spin texture is rare compared to that of the skyrmion crystal with a noncoplanar spin texture. 
Here we theoretically report the stabilization mechanisms of the bubble crystal in tetragonal magnets. 
By performing numerical calculations based on an efficient steepest descent method for an effective spin model with magnetic anisotropy and multiple spin interactions in momentum space on a two-dimensional square lattice, we construct magnetic field--temperature phase diagrams for various sets of model parameters. 
We find that the bubble crystal is stabilized at finite temperatures near the skyrmion crystal by an easy-axis anisotropic two-spin interaction. 
Through a detailed analysis, we also show that the high-harmonic wave-vector interaction and the biquadratic interaction play important roles in the stability of the bubble crystal. 
Our results indicate a close relationship between the bubble crystal and the skyrmion crystal in terms of the stabilization mechanisms, which suggests the possibility of the bubble crystal in the skyrmion-hosting materials by controlling the easy-axis magnetic anisotropy through external and/or chemical pressure. 
\end{abstract}

\maketitle

\section{Introduction}

\begin{figure}[tb!]
\begin{center}
\includegraphics[width=0.8\hsize]{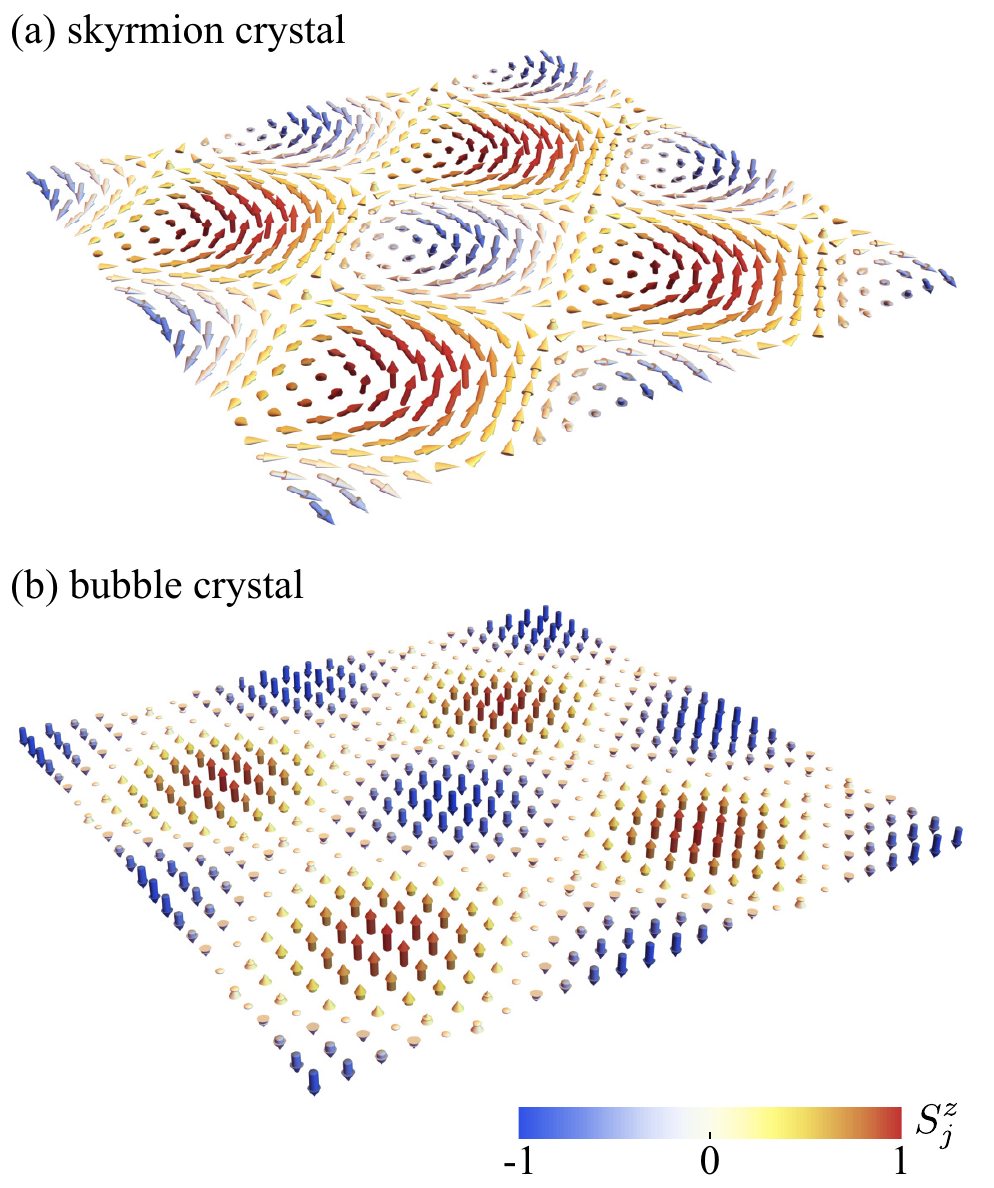} 
\caption{
\label{fig: ponti} 
Schematic spin configurations of (a) the skyrmion crystal (SkX) with the noncoplanar spin texture and (b) the bubble crystal with the collinear spin texture in a tetragonal system. 
The arrows represent the direction of the spin moments and their color shows the $z$-spin component. 
}
\end{center}
\end{figure}

Multiple-$Q$ states, which are characterized by a superposition of multiple spin density waves, have attracted great interest for years~\cite{Bak_PhysRevLett.40.800, Shapiro_PhysRevLett.43.1748, rossler2006spontaneous, Binz_PhysRevLett.96.207202, Binz_PhysRevB.74.214408, Binz2008, Yi_PhysRevB.80.054416,batista2016frustration}. 
They often manifest themselves not only in complex spin textures including noncollinear and noncoplanar ones but also in unconventional charge/spin transports. 
For example, a magnetic skyrmion crystal (SkX) consisting of multiple spiral waves in Fig.~\ref{fig: ponti}(a) leads to the topological Hall effect~\cite{nagaosa2013topological, Lee_PhysRevLett.102.186601, Neubauer_PhysRevLett.102.186602, Hamamoto_PhysRevB.92.115417} and a vortex crystal consisting of multiple sinusoidal waves leads to the nonreciprocal transport~\cite{hayami2021phase} owing to their emergent magnetic fields acting on conduction electrons. 
Since the emergent magnetic field relies on neither net magnetization nor relativistic spin--orbit coupling and its magnitude reaches $10^3$--$10^4$~T~\cite{Ohgushi_PhysRevB.62.R6065}, the use of multiple-$Q$ states for spintronics applications has been also extensively studied~\cite{zhang2020skyrmion}. 

Exploring such multiple-$Q$ states is one of the challenges in both experiments and theory. 
Especially, it is important to understand a fundamental essence to realize the multiple-$Q$ states. 
In this context, centrosymmetric tetragonal compounds, such as GdRu$_2$Si$_2$~\cite{khanh2020nanometric, Yasui2020imaging, khanh2022zoology, Matsuyama_PhysRevB.107.104421} and EuAl$_4$~\cite{Shang_PhysRevB.103.L020405, shimomura2019lattice, kaneko2021charge, Zhu_PhysRevB.105.014423,takagi2022square, Meier_PhysRevB.106.094421, Gen_PhysRevB.107.L020410, hayami2023orthorhombic}, are prototypical ones to examine the origin of the multiple-$Q$ states, since they exhibit a variety of multiple-$Q$ states despite their simple lattice structures: the square (rhombic) SkXs with (without) fourfold rotational symmetry, double-$Q$ vortex crystal, and so on. 
Their stabilization mechanisms are narrowed down owing to the simple lattice structure without geometrical frustration and the Dzyaloshinskii-Moriya interaction~\cite{dzyaloshinsky1958thermodynamic,moriya1960anisotropic}: multiple-spin interaction and high-harmonic interactions by itinerant frustration~\cite{Hayami_PhysRevB.105.174437, hayami2021topological}, dipolar interaction~\cite{Utesov_PhysRevB.103.064414}, bond-dependent anisotropic interaction~\cite{Yi_PhysRevB.80.054416, Wang_PhysRevB.103.104408}, and their competition~\cite{Hayami_PhysRevB.103.024439, Hayami_PhysRevB.105.104428} based on microscopic and phenomenological analyses~\cite{Christensen_PhysRevX.8.041022, Huang_PhysRevResearch.5.013125}. 
Especially, an effective spin model incorporating the multiple-spin interaction, high-harmonic interaction, and anisotropic two-spin interaction can qualitatively reproduce the magnetic field--temperature phase diagram in GdRu$_2$Si$_2$~\cite{khanh2022zoology, hayami2023widely} and the ground-state phase diagram in EuAl$_4$~\cite{takagi2022square, hayami2023orthorhombic}. 
Since multiple-$Q$ instabilities have been implied in similar centrosymmetric tetragonal magnets like EuGa$_4$~\cite{zhang2022giant, Zhu_PhysRevB.105.014423, Pakhira_PhysRevB.107.024421}, Eu(Ga$_{1-x}$Al$_{x})_4$~\cite{Moya_PhysRevMaterials.6.074201, moya2023real}, Mn$_{2-x}$Zn$_x$Sb~\cite{Nabi_PhysRevB.104.174419}, and PrMn$_2$Ge$_2$~\cite{song2022critical}, there is a chance that unknown multiple-$Q$ states appear in these relevant materials. 
To stimulate experimental identification, a further theoretical investigation based on microscopic model calculations is desired.

In the present study, we focus on the stabilization conditions of a magnetic bubble crystal, which corresponds to another multiple-$Q$ state formed by a collinear-type superposition of multiple sinusoidal waves, as schematically shown in Fig.~\ref{fig: ponti}(b). 
Although it has been found in magnets with strong magnetic anisotropy~\cite{lin1973bubble, Garel_PhysRevB.26.325, takao1983study}, its candidate materials have been still limited~\cite{seo2021spin}. 
Meanwhile, it was shown that the emergence of the bubble crystal is closely related to that of the SkX in the hexagonal magnetic insulators~\cite{Hayami_PhysRevB.93.184413, Hayami_PhysRevB.103.224418} and metals~\cite{hayami2020multiple, Hayami_10.1088/1367-2630/ac3683}; the synergy between a larger easy-axis two-spin or single-ion magnetic anisotropy and an external magnetic field tends to stabilize the bubble crystal. 
Thus, one can expect that a similar instability toward the bubble crystal occurs in the skyrmion-hosting tetragonal magnets by increasing the easy-axis two-spin magnetic anisotropy under external pressure and/or chemical substitution. 
The systematic investigation of the magnetic field--temperature phase diagram in such a situation is a reference to encourage searching for the materials hosting the bubble crystal. 

To get an insight into the stabilization of the bubble crystal and its relevance with that of the SkX in centrosymmetric tetragonal magnets, we construct the magnetic field--temperature phase diagram of an effective spin model with momentum-resolved interaction on a two-dimensional square lattice by performing an efficient steepest descent method, which enables us to obtain thermodynamic phases with small computational cost~\cite{hayami2023widely}. 
By systematically changing the degree of the easy-axis two-spin magnetic anisotropy, we find two additional key ingredients for the square-type bubble crystal: One is the high-harmonic wave-vector interaction and the other is the biquadratic interaction. 
We show the emergence of the finite-temperature phase transition from the bubble crystal to the SkX by lowering the temperature in both mechanisms when the strength of the easy-axis anisotropic two-spin interaction is moderate. 
Furthermore, we show that the bubble crystal is realized in the ground state as the degree of the easy-axis two-spin anisotropy increases. 
From the obtained phase diagrams, we discuss the conditions of the bubble crystal in tetragonal magnets, which will be useful for future exploration in experiments. 

The rest of this paper is organized as follows. 
In Sec.~\ref{sec: Model and method}, we describe the effective spin model on the square lattice and briefly introduce the numerical method to derive thermodynamic states. 
We discuss the stability of the bubble crystal by constructing the magnetic field--temperature phase diagrams in Sec.~\ref{sec: Results}. 
We show the behavior of the bubble crystal under the high-harmonic wave-vector interaction and biquadratic interaction in Sec.~\ref{sec: Higher-harmonic wave-vector interaction} and Sec~\ref{sec: Biquadratic interaction}, respectively. 
Finally, a summary is presented in Sec.~\ref{sec: Summary}. 
We show the supplementary data about the phase diagram in the presence of the high-harmonic wave-vector interaction in Appendix~\ref{appendix}.

\section{Model and method}
\label{sec: Model and method}

\subsection{Model}
\label{sec: Model}

\begin{figure}[tb!]
\begin{center}
\includegraphics[width=1.0\hsize]{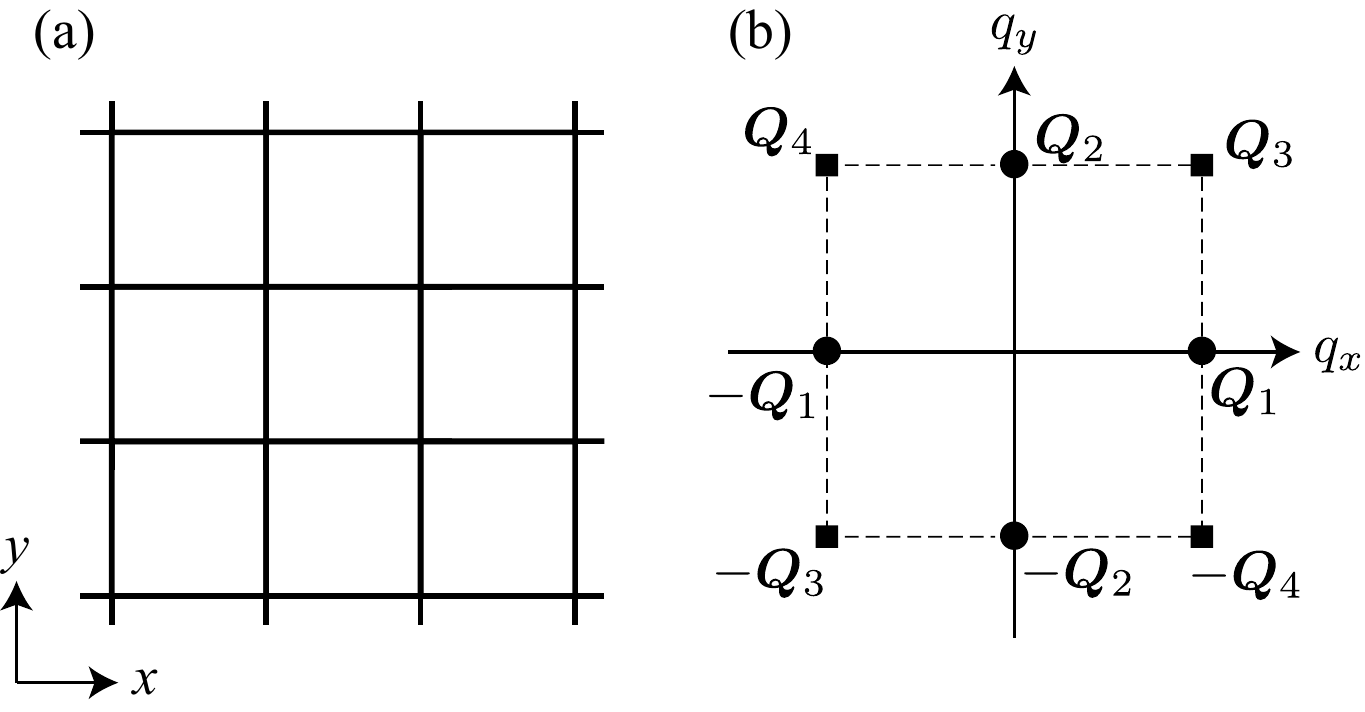} 
\caption{
\label{fig: model} 
(a) Square lattice structure in real space. 
(b) The representative wave vectors of the model in Eq.~(\ref{eq: Ham}) in momentum space. 
The interactions at $\bm{Q}_1=(Q,0)$ and $\bm{Q}_2=(0,Q)$ [$\bm{Q}_3=(Q,Q)$ and $\bm{Q}_4=(-Q,Q)$] denoted by the circles (squares) represent the dominant (sub-dominant) interactions.
}
\end{center}
\end{figure}

We consider the effective spin model, which is obtained by perturbatively expanding the Kondo lattice model in terms of the spin-charge coupling between itinerant electron and classical localized spins~\cite{Hayami_PhysRevB.95.224424}. 
The model is given by 
\begin{align}
\label{eq: Ham}
\mathcal{H}=  &-J \sum_{\nu,\alpha,\beta}\Gamma^{\alpha\beta}_{\bm{Q}_\nu}S^{\alpha}_{\bm{Q}_\nu}S^{\beta}_{-\bm{Q}_\nu} \nonumber \\
&+\frac{K}{N} \sum_{\nu}\left(\sum_{\alpha,\beta}\Gamma^{\alpha\beta}_{\bm{Q}_\nu}S^{\alpha}_{\bm{Q}_\nu}S^{\beta}_{-\bm{Q}_\nu}\right)^2  
-  H \sum_{j}   S^z_{j},
\end{align}
where $S^\alpha_{\bm{Q}_\nu}$ in the first and second terms represents the momentum representation of the spin with the wave vector $\pm\bm{Q}_1, \pm\bm{Q}_2, \cdots, \pm\bm{Q}_{N_{\bm{Q}}}$ and the spin component $\alpha=x,y,z$; $N_{\bm{Q}}$ is the number of channels in the momentum-resolved interactions. 
$S^z_{j}$ in the third term represents the $z$ component of the classical localized spin at site $j$ with the unit spin length $|\bm{S}_{j}|=1$. 
$S^\alpha_{j}$ and $S^\alpha_{\bm{Q}_\nu}$ are related by the Fourier transformation with each other; $S^\alpha_{\bm Q} =  \sum_j S^\alpha_j e^{-i {\bm Q}\cdot{\bm r}_j}/\sqrt{N}$; $N$ is the total number of spins and $\bm{r}_j$ is the position vector at site $j$. 

The Hamiltonian in Eq.~(\ref{eq: Ham}) consists of three terms: the bilinear spin interaction, biquadratic spin interaction, and Zeeman coupling under an external magnetic field along the $z$ direction with the coupling constants, $J>0$, $K \geq 0$, and $H \geq 0$, respectively. 
We set $J=1$ as the energy unit. 
The first term originates from the second-order contribution with respect to the Kondo coupling, while the second term is from the fourth-order one in the perturbation scheme~\cite{Akagi_PhysRevLett.108.096401, Hayami_PhysRevB.90.060402, Ozawa_doi:10.7566/JPSJ.85.103703, Hayami_PhysRevB.95.224424}; we neglect other four-spin interactions between different $\bm{Q}_\nu$ for simplicity. 
$\Gamma^{\alpha\beta}_{\bm{Q}_\nu}$ in the first and second terms in Eq.~(\ref{eq: Ham}) represents the momentum-dependent anisotropic form factor that arises from the relativistic spin--orbit coupling and dipolar interaction. 

Nonzero components of $\Gamma^{\alpha\beta}_{\bm{Q}_\nu}$ depend on the symmetry of the lattice structure and the wave vector $\bm{Q}_\nu$~\cite{yambe2022effective}. 
We consider the two-dimensional square lattice, as shown in Fig.~\ref{fig: model}(a) by setting $\bm{r}_j=(r^x_j, r^y_j)$ with the lattice constant as unity and integers $r^x_j$ and $r^y_j$, where site symmetry in each lattice site is $D_{\rm 4h}$. 
By supposing the situation where the Fermi surface nesting at fourfold symmetric $\pm \bm{Q}_1=\pm(Q,0)$ and $ \pm \bm{Q}_2=\pm(0,Q)$ with $Q=\pi/3$ is important, the dominant interactions are expressed by those at $\bm{Q}_1$ and $\bm{Q}_2$ channels.
In such a situation, $\Gamma^{\alpha\beta}_{\bm{Q}_\nu}$ has only diagonal components regarding spins $(\alpha = \beta)$, which is represented by 
\begin{align}
&\bm{\Gamma}_{\bm{Q}_1}\equiv (\Gamma^{xx}_{\bm{Q}_1}, \Gamma^{yy}_{\bm{Q}_1}, \Gamma^{zz}_{\bm{Q}_1})=(\Gamma_x, \Gamma_y, \Gamma_z),  \nonumber \\
&\bm{\Gamma}_{\bm{Q}_2}\equiv (\Gamma^{xx}_{\bm{Q}_2}, \Gamma^{yy}_{\bm{Q}_2}, \Gamma^{zz}_{\bm{Q}_2})=(\Gamma_y, \Gamma_x, \Gamma_z), 
\end{align}
otherwise $\Gamma^{\alpha\beta}_{\bm{Q}_\nu}=0$. 
We set $\Gamma_x= 0.855 \kappa$, $\Gamma_y=  0.95 \kappa$, and $\Gamma_z=1$, where $ \kappa \geq 0$ represents the parameter for the easy-axis anisotropic two-spin interaction that originates from the spin--orbit coupling. 
It is noted that $\Gamma_x=\Gamma_y=\Gamma_z$ in the absence of the spin--orbit coupling; $\kappa$ tends to deviate from the isotropic relation for a larger spin--orbit coupling. 
The result for $\kappa=1$ reproduces the previous result to investigate the instability toward the SkX~\cite{khanh2022zoology, hayami2023widely}; 
$\kappa<1$ ($\kappa>1$) enhances (suppresses) the degree of the easy-axis anisotropic two-spin interaction. 
It is noted that finite-temperature phase transitions are allowed in the two-dimensional system in the presence of the anisotropy. 

The model in Eq.~(\ref{eq: Ham}) exhibits a variety of multiple-$Q$ states with changing $K$, $H$, and the temperature $T$, as demonstrated for $\kappa=1$~\cite{Hayami_PhysRevB.103.024439, hayami2023widely}. 
Especially, the double-$Q$ SkX consisting of a superposition of two spiral waves with $\bm{Q}_1$ and $\bm{Q}_2$ is stabilized from zero to finite temperatures for nonzero $K$ and $H$. 
With this in mind, we examine the instability toward the double-$Q$ bubble crystal, which is characterized by a superposition of two sinusoidal waves with $\bm{Q}_1$ and $\bm{Q}_2$, while changing $\kappa$. 

In addition, we consider another scenario to realize the bubble crystal by taking into account the sub-dominant contributions from the high-harmonic wave vectors $\pm \bm{Q}_3=\pm (\bm{Q}_1+\bm{Q}_2)$ and $\pm\bm{Q}_4=\pm(-\bm{Q}_1+\bm{Q}_2)$, which is expressed as
\begin{align}
\bm{\Gamma}_{\bm{Q}_3}=\bm{\Gamma}_{\bm{Q}_4}=(\kappa \Gamma', \kappa \Gamma', \Gamma').
\end{align} 
The introduction of $\bm{\Gamma}_{\bm{Q}_3}$ and $\bm{\Gamma}_{\bm{Q}_4}$ in addition to $\bm{\Gamma}_{\bm{Q}_1}$ and $\bm{\Gamma}_{\bm{Q}_2}$ favors the double-$Q$ superposition of $\bm{Q}_1$ and $\bm{Q}_2$ compared to the individual single-$Q$ spiral of $\bm{Q}_1$ or $\bm{Q}_2$ even when $K=0$~\cite{hayami2022multiple, Hayami_PhysRevB.105.174437, hayami2023widely}. 
The relation among $\bm{Q}_1$--$\bm{Q}_4$ is presented in Fig.~\ref{fig: model}(b).
We here neglect the inplane anisotropic component $\Gamma^{xy}_{\bm{Q}_3}=\Gamma^{yx}_{\bm{Q}_3}$ and $\Gamma^{xy}_{\bm{Q}_4}=\Gamma^{yx}_{\bm{Q}_4}$ for simplicity by assuming that the magnitude of $\Gamma'$ ($\kappa  \Gamma'$) is smaller than that of $\Gamma_{z}$ ($\Gamma_{x,y}$). 
For $\kappa=1$ and $K=0$, the instability toward the SkX occurs for $\Gamma' \gtrsim 0.1$~\cite{hayami2023widely}.
We discuss the possibility of the bubble crystal when introducing $\kappa \neq 1$ for the model with nonzero $\Gamma'$.

\subsection{Method}
\label{sec: Method}

The thermodynamics and optimal spin configurations of the model in Eq.~(\ref{eq: Ham}) are investigated by using the steepest descent method~\cite{Kato_PhysRevB.105.174413,hayami2023widely}, in which we perform numerical optimizations of the free energy using a JAX-based~\cite{JAX2018} library, Optax~\cite{OpTax2020}. 
This method provides a numerically exact solution in the thermodynamic limit at any temperature, which has been used for similar models to investigate various multiple-$Q$ states~\cite{Kato_PhysRevB.105.174413, hayami2023widely, Kato_PhysRevB.107.094437}. 

To identify magnetic phases for each $(\kappa, K, \Gamma', H)$, we calculate the $\bm{Q}_\nu$ components of the magnetic moments $m^{\alpha}_{\bm{Q}_\nu}$ and scalar chirality $\chi^{\rm sc}$, which are given by 
\begin{align}
m^\alpha_{\bm{Q}_\nu} &= \frac{1}{N}\sqrt{
\sum_{j,j'}
\left\langle 
S^\alpha_j S^\alpha_{j'}
\right\rangle
e^{i \bm{Q}_\nu \cdot (\bm{r}_j - \bm{r}_{j'})}}, \\
\chi^{\rm sc}&= \frac{1}{2 N} 
\sum_{j}
\sum_{\delta,\delta'= \pm1}
\delta \delta'
\left\langle \bm{S}_{j} \cdot (\bm{S}_{j+\delta\hat{x}} \times \bm{S}_{j+\delta'\hat{y}}) \right\rangle, 
\end{align}
where $\hat{x}$ ($\hat{y}$) represents a translation by the lattice constant along the $x$ ($y$) direction. 
We also calculate the $z$ component of the total magnetization as 
\begin{align}
M^z&= \frac{1}{N} \sum_{j} \langle S^z_{j} \rangle. 
\end{align}

\section{Results}
\label{sec: Results}

We discuss the instability toward the bubble crystal on the square lattice by focusing on two mechanisms: the higher-harmonic wave-vector interaction in Sec.~\ref{sec: Higher-harmonic wave-vector interaction} and the biquadratic interaction in Sec.~\ref{sec: Biquadratic interaction}. 
We show magnetic field--temperature phase diagrams for the different easy-axis magnetic anisotropic two-spin interactions in each case.

\subsection{Higher-harmonic wave-vector interaction}
\label{sec: Higher-harmonic wave-vector interaction}

\begin{figure*}[tb!]
\begin{center}
\includegraphics[width=1.0\hsize]{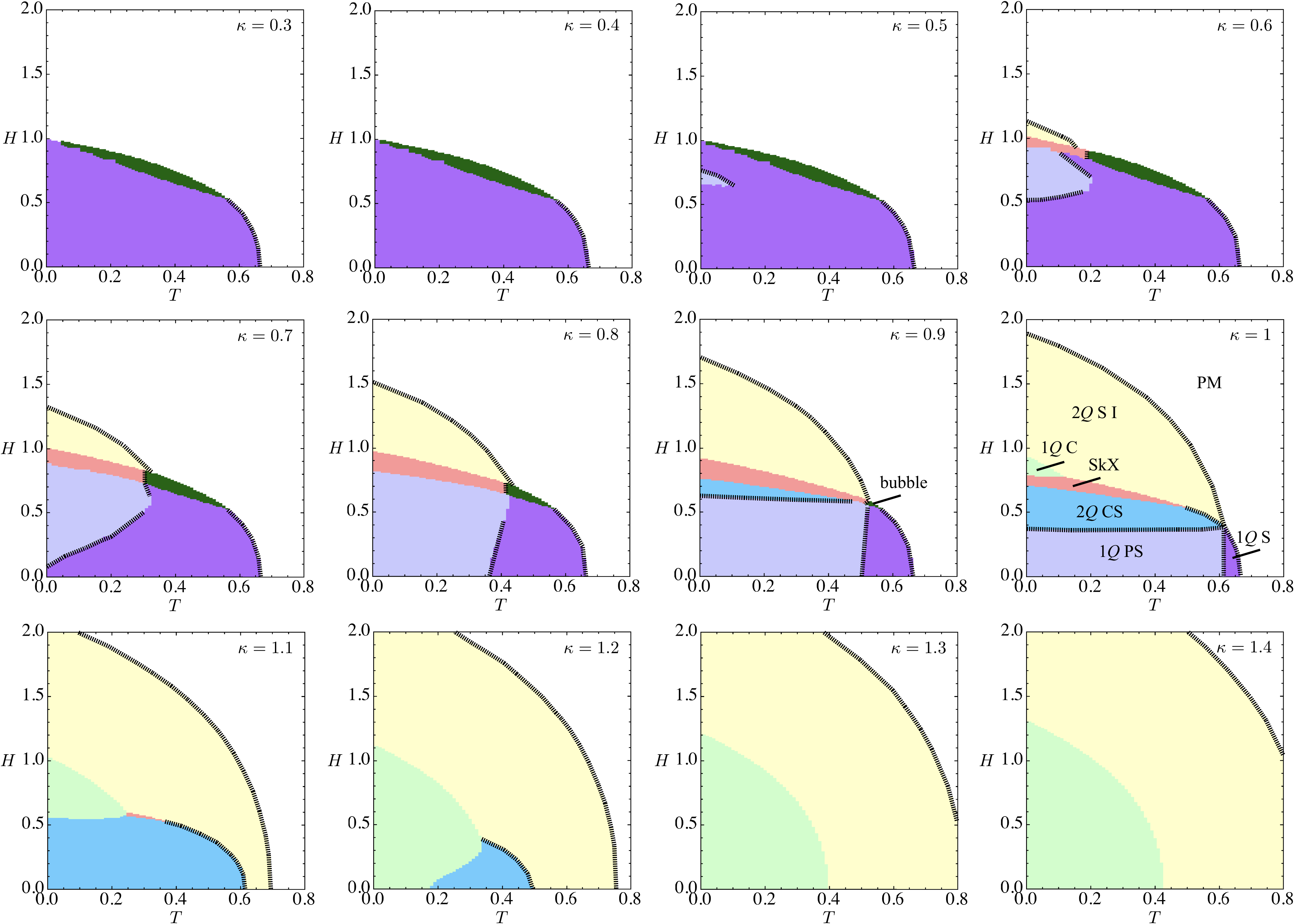} 
\caption{
\label{fig: PD_HH} 
Magnetic field ($H$)--temperature ($T$) phase diagrams of the model in Eq.~(\ref{eq: Ham}) at $\Gamma'=0.3$ and $K=0$ with changing $\kappa$ for $0.3 \leq \kappa \leq 1.4$ by $\Delta \kappa =0.1$.  
$1Q$ and $2Q$ represent the single-$Q$ and double-$Q$ states, respectively. 
PS, C, S, CS, SkX, and PM stand for proper-screw, conical, sinusoidal, chiral stripe, skyrmion crystal, and paramagnetic state, respectively. 
The dashed lines represent the second-order phase transitions.
The phase diagram at $\kappa=1$ is the same as that in Ref.~\cite{hayami2023widely}. 
}
\end{center}
\end{figure*}

\begin{figure}[t!]
\begin{center}
\includegraphics[width=1.0\hsize]{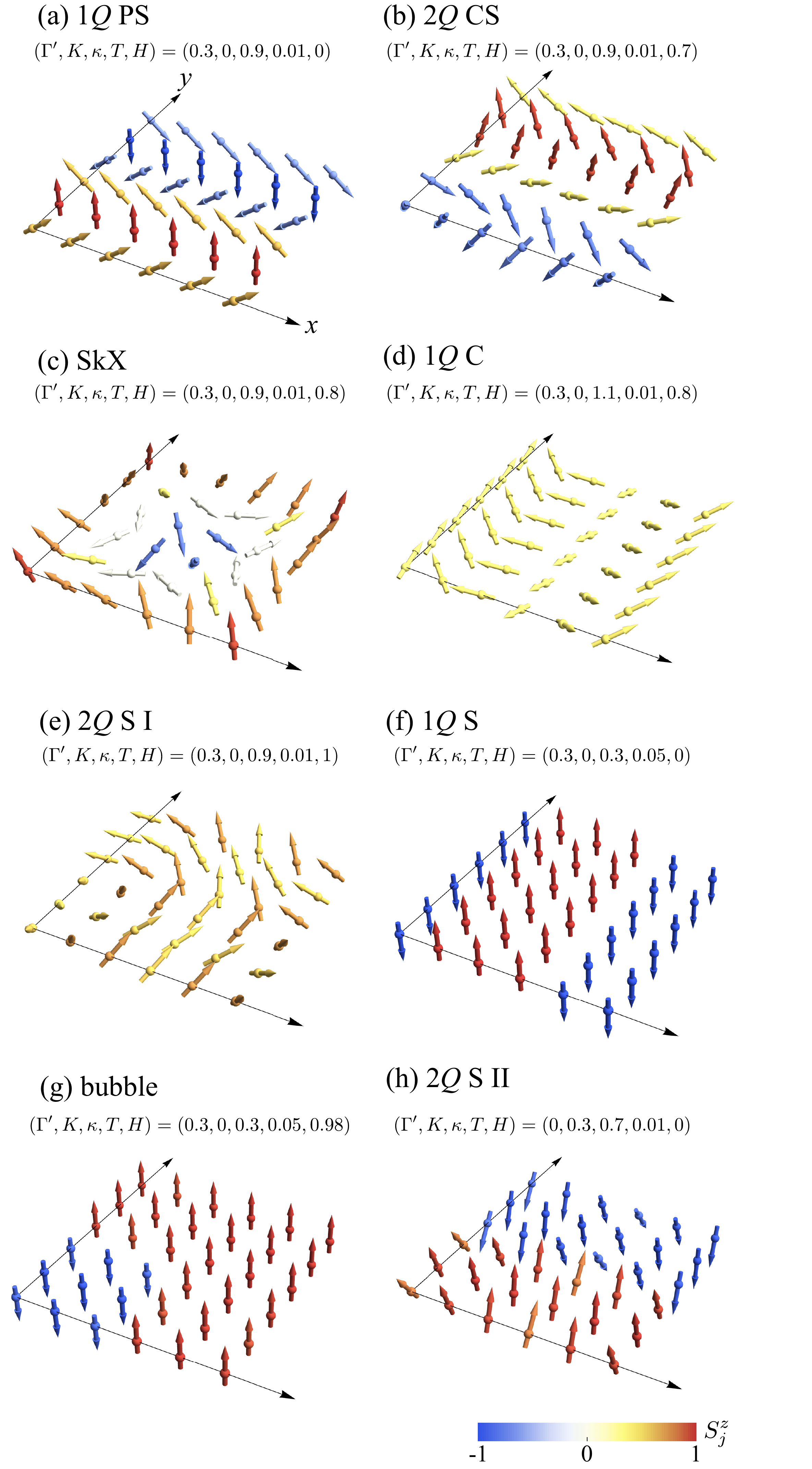} 
\caption{
\label{fig: spin} 
The spin configurations of (a) the 1$Q$ PS, (b) 2$Q$ CS, (c) SkX, (d) 1$Q$ C, (e) 2$Q$ S I, (f) 1$Q$ S, (g) bubble crystal, and (h) 2$Q$ S II states in the $6 \times 6$ magnetic unit cell, which appear in Figs.~\ref{fig: PD_HH} and \ref{fig: PD_K}. 
The arrows represent the direction of the spin moments and their color shows the $z$-spin component. 
The model parameters $(\Gamma', K, \kappa, T, H)$ are shown in each figure. 
}
\end{center}
\end{figure}

\begin{figure}[t!]
\begin{center}
\includegraphics[width=1.0\hsize]{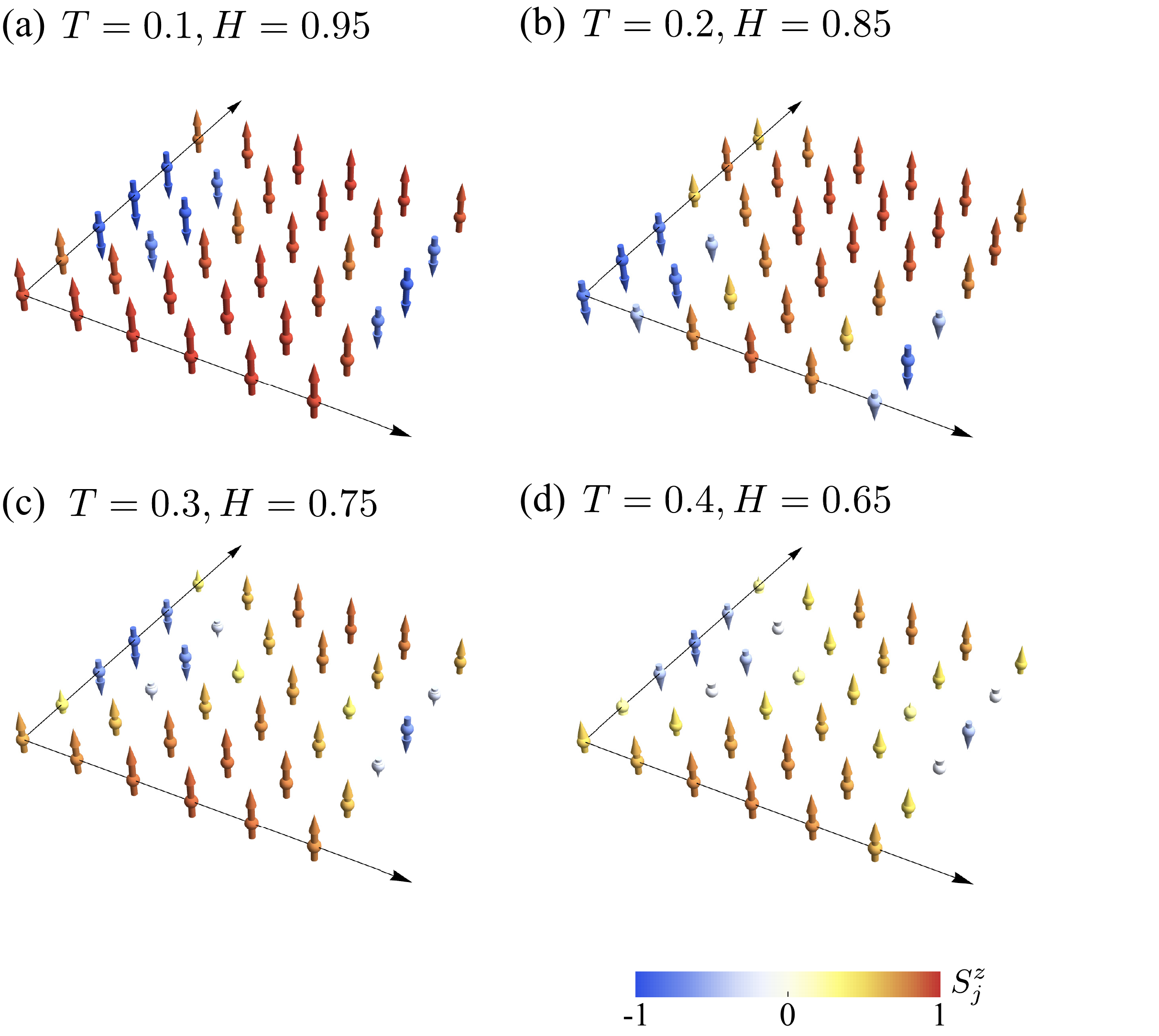} 
\caption{
\label{fig: spin_bubble} 
The spin configurations of the bubble crystal at (a) $T=0.1$ and $H=0.95$, (b) $T=0.2$ and $H=0.85$, (c) $T=0.3$ and $H=0.75$, and (d) $T=0.4$ and $H=0.65$ for $\Gamma'=0.3$ and $K=0$ in the $6 \times 6$ magnetic unit cell. 
The arrows represent the direction of the spin moments and their color shows the $z$-spin component. 
}
\end{center}
\end{figure}

\begin{figure}[t!]
\begin{center}
\includegraphics[width=0.76\hsize]{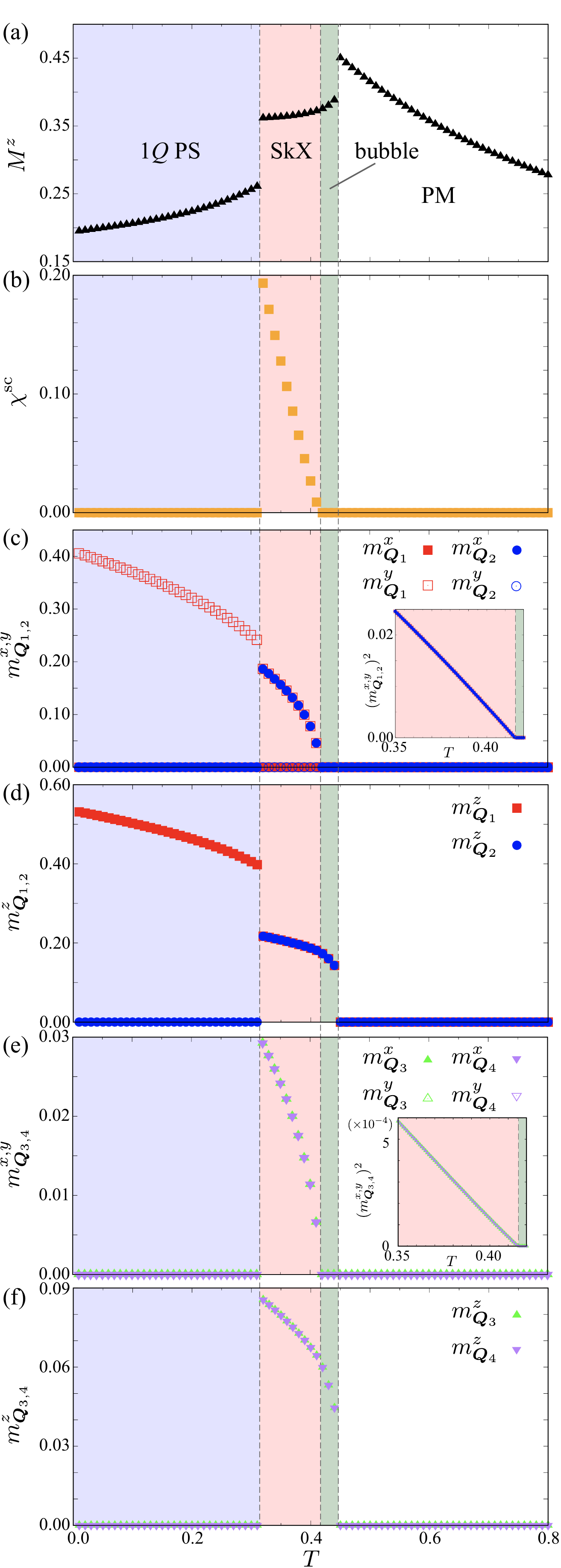} 
\caption{
\label{fig: Mq_HH} 
$T$ dependence of (a) the magnetization $M^z$, (b) the scalar chirality $\chi^{\rm sc}$, (c) $m^{x,y}_{\bm{Q}_{1,2}}$, (d) $m^{z}_{\bm{Q}_{1,2}}$, (e) $m^{x,y}_{\bm{Q}_{3,4}}$, and (f) $m^{z}_{\bm{Q}_{3,4}}$ at $\Gamma'=0.3$, $K=0$, $\kappa=0.8$, and $H=0.7$. 
The inset of (c) [(e)] shows $(m^{y}_{\bm{Q}_{1}})^2$ and $(m^{x}_{\bm{Q}_{2}})^2$ [$(m^{y}_{\bm{Q}_{3}})^2=(m^{x}_{\bm{Q}_{3}})^2$ and $(m^{x}_{\bm{Q}_{4}})^2=(m^{y}_{\bm{Q}_{4}})^2$] in the vicinity of the phase boundary between the SkX and bubble crystal.
The vertical dashed lines represent the phase boundaries. 
The blue, red, green, and white regions represent the 1$Q$ PS, SkX, bubble crystal, and PM states, respectively. 
}
\end{center}
\end{figure}

First, we discuss the result in the presence of the high-harmonic wave-vector interaction; we set $K=0$ in this section. 
Figure~\ref{fig: PD_HH} shows the magnetic field($H$)--temperature($T$) phase diagrams at $\Gamma'=0.3$ with changing $\kappa$ from $0.3$ to $1.4$ by $0.1$, which is obtained by the steepest descent method in Sec.~\ref{sec: Method}.  

\subsubsection{Case of $\kappa=1$}
The result at $\kappa=1$ has been discussed in Ref.~\cite{hayami2023widely}, where six phases including the SkX appear but the bubble crystal does not appear in the phase diagram. 
In the low-$T$ and low-$H$ region, the single-$Q$ proper-screw spiral (1$Q$ PS) state is stabilized, whose ordering vector is $\bm{Q}_{1,2}$ not $\bm{Q}_{3,4}$ owing to $\Gamma'<\Gamma_{x,y,z}$ and spiral plane is perpendicular to $\bm{Q}_{1,2}$ owing to $\Gamma_x < \Gamma_y < \Gamma_z$. 
The real-space spin configuration of the 1$Q$ PS state is presented in Fig.~\ref{fig: spin}(a). 
When $H$ increases, the 1$Q$ PS state continuously changes into the double-$Q$ chiral stripe (2$Q$ CS) state. 
This state is characterized by a superposition of the proper-screw spiral wave at $\bm{Q}_1$ ($\bm{Q}_2$) and the sinusoidal wave at $\bm{Q}_2$ ($\bm{Q}_1$), i.e., $m^y_{\bm{Q}_1}, m^z_{\bm{Q}_1}, m^x_{\bm{Q}_2} \neq 0$ ($m^y_{\bm{Q}_1}, m^x_{\bm{Q}_2}, m^z_{\bm{Q}_2} \neq 0$)~\cite{Ozawa_doi:10.7566/JPSJ.85.103703}. 
Since the amplitude of the sinusoidal component is smaller than that of the proper-screw spiral component, the real-space spin configuration of the 2$Q$ CS state resembles that of the 1$Q$ PS state, as shown in Figs.~\ref{fig: spin}(a) and \ref{fig: spin}(b); the small sinusoidal $y$-spin oscillation along the $x$ direction can be found in the 2$Q$ CS state. 
Reflecting the double-$Q$ structure, $m^x_{\bm{Q}_3}=m^x_{\bm{Q}_4}$ (or $m^y_{\bm{Q}_3}=m^y_{\bm{Q}_4}$) is slightly induced, which is smaller than $|\bm{m}_{\bm{Q}_1}|$ and $|\bm{m}_{\bm{Q}_2}|$. 

With a further increase of $H$, the 2$Q$ CS state is replaced by the SkX through the first-order phase transition. 
The SkX consists of two proper-screw spiral waves at $\bm{Q}_1$ and $\bm{Q}_2$; $m^y_{\bm{Q}_1}=m^x_{\bm{Q}_2} \neq 0$ and $m^z_{\bm{Q}_1}=m^z_{\bm{Q}_2} \neq 0$ so that the fourfold rotational or improper rotational symmetry is preserved. 
In addition, $m^x_{\bm{Q}_3}=m^y_{\bm{Q}_3}=m^x_{\bm{Q}_4}=m^y_{\bm{Q}_4}$ and $m^z_{\bm{Q}_3}=m^z_{\bm{Q}_4}$ become nonzero owing to the superposition of the spiral waves at $\bm{Q}_1$ and $\bm{Q}_2$, which means that $\Gamma'$ assists the stabilization of the SkX. 
Indeed, the SkX phase vanishes for $\Gamma'=0$~\cite{hayami2023widely}. 
The SkX is the only phase to have nonzero scalar chirality $\chi^{\rm sc}$ in the phase diagram. 
It is noted that there is a degeneracy between the SkX with positive $\chi^{\rm sc}$ and negative one. 
Such a degeneracy is lifted by considering the anisotropic form factor in the $\bm{Q}_{3,4}$ channel; $\Gamma^{xy}_{\bm{Q}_3}>0$ ($\Gamma^{xy}_{\bm{Q}_3}<0$) leads to the anti-type (Bloch-type) SkX with positive (negative) $\chi^{\rm sc}$~\cite{Hayami_doi:10.7566/JPSJ.89.103702}. 
We present the spin configuration of the SkX in Fig.~\ref{fig: spin}(c), where the anti-type SkX emerges. 

The SkX shows a first-order phase transition to the single-$Q$ conical spiral (1$Q$ C) state with increasing $H$. 
The 1$Q$ C state is represented by the spiral modulation on the $xy$ plane with $\bm{Q}_{1,2}$, whose spin configuration is shown in Fig.~\ref{fig: spin}(d).  
The 1$Q$ C state changes into the double-$Q$ sinusoidal I (2$Q$ S I) state with increasing $H$ discontinuously. 
The 2$Q$ S I state consists of $y$($x$)-spin sinusoidal oscillations at $\bm{Q}_1$ ($\bm{Q}_2$) with the equal amplitude, i.e., $m^y_{\bm{Q}_1}=m^x_{\bm{Q}_2}$. 
The spin configuration is presented in Fig.~\ref{fig: spin}(e). 
In contrast to the 2$Q$ CS and SkX with the double-$Q$ spin modulations, no $m^{\alpha}_{\bm{Q}_3}$ and $m^{\alpha}_{\bm{Q}_4}$ are induced in the 2$Q$ S I state. 
The 2$Q$ S I state continuously turns into the paramagnetic (PM) state without the finite-$q$ modulation. 

With the increase of $T$ in the low-$H$ region, the single-$Q$ sinusoidal (1$Q$ S) state appears. 
The 1$Q$ S state has a $z$-spin oscillation without $x,y$-spin ones, i.e., $m^z_{\bm{Q}_1}\neq 0$ and $m^x_{\bm{Q}_1}=m^y_{\bm{Q}_1} = 0$ or $m^z_{\bm{Q}_2}\neq 0$ and $m^x_{\bm{Q}_2}=m^y_{\bm{Q}_2} = 0$. 
The appearance of the 1$Q$ S state is attributed to $\Gamma^z > \Gamma_{x,y}$.  

\subsubsection{Case of $\kappa<1$}

For the large easy-axis magnetic anisotropic two-spin interaction, i.e., $\kappa<1$, the bubble crystal is realized for $\kappa \lesssim 0.9$ in the intermediate-field and finite-temperature regions. 
As shown in the phase diagram at $\kappa=0.9$ in Fig.~\ref{fig: PD_HH}, the bubble crystal emerges in the competing region among the 1$Q$ PS, 1$Q$ S, SkX, and 2$Q$ S I phases. 
For smaller $\kappa$, the region where the bubble crystal is stabilized extends to a lower-$T$ region but not to a zero temperature. 
The bubble crystal is mainly expressed as a collinear superposition of two sinusoidal waves with the $z$-spin component at $\bm{Q}_1$ and $\bm{Q}_2$; $m^z_{\bm{Q}_1}$ equals to $m^z_{\bm{Q}_2}$.  
In addition, $m^z_{\bm{Q}_3}=m^z_{\bm{Q}_4}$ is also induced, although it is smaller than $m^z_{\bm{Q}_{1,2}}$. 
In the real-space spin configuration in Fig.~\ref{fig: spin}(g), a quarter of the number of spins in the magnetic unit cell point along the $-z$ direction, while the remaining spins point along the $z$ direction; the magnetization per magnetic unit cell takes 0.5 in the zero-temperature limit. 
When the effect of thermal fluctuations is considered, spin moments shrink nonuniformly, as shown in the case of $\kappa=0.3$ in Figs.~\ref{fig: spin_bubble}(a)--\ref{fig: spin_bubble}(d). 
The tendency of the nonuniform spin-length distribution becomes larger for larger $T$.

We show the $T$ dependence of physical quantities in the phase transition between the bubble crystal and the SkX at $\Gamma'=0.3$, $K=0$, $\kappa=0.8$, and $H=0.7$ in Fig.~\ref{fig: Mq_HH}. 
The data of $M^z$, $\chi^{\rm sc}$, $m^{x,y}_{\bm{Q}_{1,2}}$, $m^{z}_{\bm{Q}_{1,2}}$, $m^{x,y}_{\bm{Q}_{3,4}}$, and $m^{z}_{\bm{Q}_{3,4}}$ are shown in Figs.~\ref{fig: Mq_HH}(a)--\ref{fig: Mq_HH}(f), respectively. 
In this set of model parameters, the phase sequence is represented by the 1$Q$ PS, SkX, bubble crystal, and PM states with increasing $T$. 
The phase transition between the 1$Q$ PS and SkX and that between the bubble crystal and PM state are of the first order~\cite{comment_bubble_PM}, while the transition between the SkX and bubble crystal is of the second order; see the $T$ dependence of $M^z$ in Fig.~\ref{fig: Mq_HH}(a) for example. 
In the latter phase transition, both $m^{z}_{\bm{Q}_{1,2}}$ and $m^{z}_{\bm{Q}_{3,4}}$ remain nonzero with keeping the double-$Q$ structure, as shown in Figs.~\ref{fig: Mq_HH}(d) and \ref{fig: Mq_HH}(f). 
Meanwhile, both $m^{x,y}_{\bm{Q}_{1,2}}$ and $m^{x,y}_{\bm{Q}_{3,4}}$ in the SkX phase gradually become smaller as $T$ increases, which follow the mean-field critical exponent as $(m^{x,y}_{\bm{Q}_{1,2}})^2 \propto T_{\rm c}-T$ and $(m^{x,y}_{\bm{Q}_{3,4}})^2 \propto T_{\rm c}-T$ ($T_{\rm c}$ is the transition temperature), as shown in the inset of Figs.~\ref{fig: Mq_HH}(c) and \ref{fig: Mq_HH}(e), respectively, and vanish in the bubble crystal phase, as shown in Figs.~\ref{fig: Mq_HH}(c) and \ref{fig: Mq_HH}(e). 
Accordingly, $\chi^{\rm sc}$ vanishes in the bubble crystal phase, as shown in Fig.~\ref{fig: Mq_HH}(b). 

The emergence of the bubble crystal is attributed to the interplay between the easy-axis anisotropic two-spin interaction $\kappa$ and the high-harmonic wave-vector interaction $\Gamma'$ under the magnetic field. 
This is understood from the effective coupling in the form of $S^z_{\bm{Q}_1}S^z_{\bm{Q}_2}S^z_{-\bm{Q}_3} S^z_{\bm{0}}$ and $S^z_{-\bm{Q}_1} S^z_{\bm{Q}_2}S^z_{-\bm{Q}_4} S^z_{\bm{0}}$ appearing in the free energy, where the magnitude of $S^z_{\bm{Q}_3}$ and $S^z_{\bm{Q}_4}$ depends on $\Gamma'$. 
Since larger $\Gamma'$ tends to make $S^z_{\bm{Q}_3}$ and $S^z_{\bm{Q}_4}$ larger, it results in the stabilization of the bubble crystal, as will be discussed in more detail in Sec.~\ref{sec: Case of other Gamma}. 
The above effective coupling indicates that uniform magnetization also plays an important role in stabilizing the bubble crystal, which might be a reason why the bubble crystal appears only for nonzero $H$ in Fig.~\ref{fig: PD_HH}. 

Intriguingly, the bubble crystal appears next to the SkX upon increasing $T$ for the moderate easy-axis anisotropic two-spin interaction $0.6 \lesssim \kappa \lesssim 0.9$, as shown in Fig.~\ref{fig: PD_HH}. 
This is naturally understood from the difference in the spin configurations between the bubble crystal and SkX. 
The spin configuration for these phases is commonly well approximated by 
\begin{align}
\label{eq: SkX}
\bm{S}_j \propto 
\left(
    \begin{array}{c}
    a_{xy}(\cos \mathcal{Q}_1+  \cos \mathcal{Q}_2) \\
    a_{xy}(\cos \mathcal{Q}_1 -  \cos \mathcal{Q}_2) \\
    a_z (\sin \mathcal{Q}_1 +\sin \mathcal{Q}_2)+ \tilde{M}_z
          \end{array}
  \right)^{\rm T}, 
\end{align}
where T is the transpose of the vector and $\mathcal{Q}_\nu=\bm{Q}_\nu \cdot \bm{r}_{j}+ \phi_\nu$; $\phi_\nu$ is the phase degree of freedom in the spiral wave and is chosen so that the skyrmion core is located at the center of the square plaquette so as to satisfy $S^z_j \neq -1$ for all $j$, as shown in Fig.~\ref{fig: spin}(c)~\cite{Hayami_PhysRevResearch.3.043158}. 
$a_{xy}$, $a_z$, and $\tilde{M}_z$ are variational parameters, which are related to $m^{x,y}_{\bm{Q}_{1,2}}$, $m^{z}_{\bm{Q}_{1,2}}$, and $M^z$, respectively. 
These parameters are determined by the model parameters and $T$. 
The spin configuration with nonzero $a_{xy}$, $a_{z}$, and $\tilde{M}_z$ corresponds to the SkX, while that with nonzero $a_{z}$ and $\tilde{M}_z$ but $a_{xy}=0$ corresponds to the bubble crystal. 
Since the energy scale of the $z$-spin component is larger than that of the $xy$-spin component owing to $\Gamma_z > \Gamma_{x,y}$, the situation where only the $z$-spin component is ordered and exhibits the double-$Q$ structure with $a_z \neq 0$ and $a_{xy}=0$ can occur when the temperature is lowered from the high-temperature PM state, which means the appearance of the bubble crystal in the high-$T$ region. 
As $T$ decreases, the $xy$-spin component additionally shows the double-$Q$ structure so that the phase turns into the SkX with $a_z \neq 0$ and $a_{xy} \neq 0$, as shown in Figs.~\ref{fig: Mq_HH}(c) and \ref{fig: Mq_HH}(e).

Finally, we discuss the behaviors of the other phases in the phase diagram in Fig.~\ref{fig: PD_HH} with changing $\kappa$. 
The states with nonzero $m^z_{\bm{Q}_\nu}$, i.e., 1$Q$ PS, 2$Q$ CS, SkX, and 1$Q$ S states, tend to be stabilized while those without $m^z_{\bm{Q}_\nu}$, i.e., $1Q$ C and $2Q$ S I, tend to be destabilized as $\kappa$ decreases from $\kappa=1$. 
As shown in the phase diagram in Fig.~\ref{fig: PD_HH}, the 1$Q$ C state vanishes at $\kappa=0.9$ and the 2$Q$ S I state vanishes at $\kappa=0.5$. 
In the low-field region, the 2$Q$ CS state is replaced by the 1$Q$ PS as shown in the phase diagrams for $\kappa=0.8$--$1$ and the 1$Q$ PS state is replaced by the 1$Q$ S state as shown in the phase diagrams for $\kappa=0.4$--$1$. 
This is because the 1$Q$ S state has a more component of $m^z_{\bm{Q}_\nu}$ compared to the other phases. 
It is noted that the spin configuration close to the high-field phase boundary in the 1$Q$ S state is characterized by the up-up-up-up-down-down structure instead of the up-up-up-down-down-down structure in Fig.~\ref{fig: spin}(f). 
In the intermediate-field region, the SkX is replaced by the bubble crystal, as discussed above.

\subsubsection{Case of $\kappa>1$}

For $\kappa>1$, the bubble crystal does not appear in contrast to the situation for $\kappa<1$, as shown in Fig.~\ref{fig: PD_HH}, which clearly indicates that the easy-axis anisotropic two-spin interaction plays an important role in stabilizing the bubble crystal. 
For the other phases, their stability region tends to be larger (smaller) as the contribution of the $xy$-spin component in the spin configuration becomes larger (smaller). 
In the low-field region, the 1$Q$ PS and 1$Q$ S states are replaced with the 2$Q$ CS state owing to the energy gain arising from the sinusoidal inplane-spin component in the latter state. 
For larger $\kappa \gtrsim 1.3$, the 2$Q$ CS state vanishes. 
In the intermediate-field region, the region of the 1$Q$ C state is extended and that of the SkX is shrunk for $\kappa=1.1$ and vanishes for $\kappa=1.2$. 
Since the relative amplitude relation of $\Gamma_{x,y,z}$ is switched from $\Gamma_z>\Gamma_y > \Gamma_x$ to $\Gamma_y>\Gamma_z > \Gamma_x$ at $\kappa \simeq 1.053$, and to $\Gamma_y> \Gamma_x >\Gamma_z$ at $\kappa \simeq 1.170$, the result indicates that the relation of $\Gamma_z > \Gamma_x$ is important in stabilizing the SkX.
This is naturally understood from the fact that the SkX consists of two vertical spiral waves instead of the inplane spiral waves, the latter of which is realized when $\Gamma_y> \Gamma_x >\Gamma_z$. 
In the high-field region, the 2$Q$ S I state is more stabilized for $\kappa>1$, whose region is extended to that for $H=0$ at finite temperatures, as also found in frustrated localized spin models~\cite{Utesov_PhysRevB.103.064414, Wang_PhysRevB.103.104408}.

\subsubsection{Case of other $\Gamma'$}
\label{sec: Case of other Gamma}

\begin{figure}[t!]
\begin{center}
\includegraphics[width=1.0\hsize]{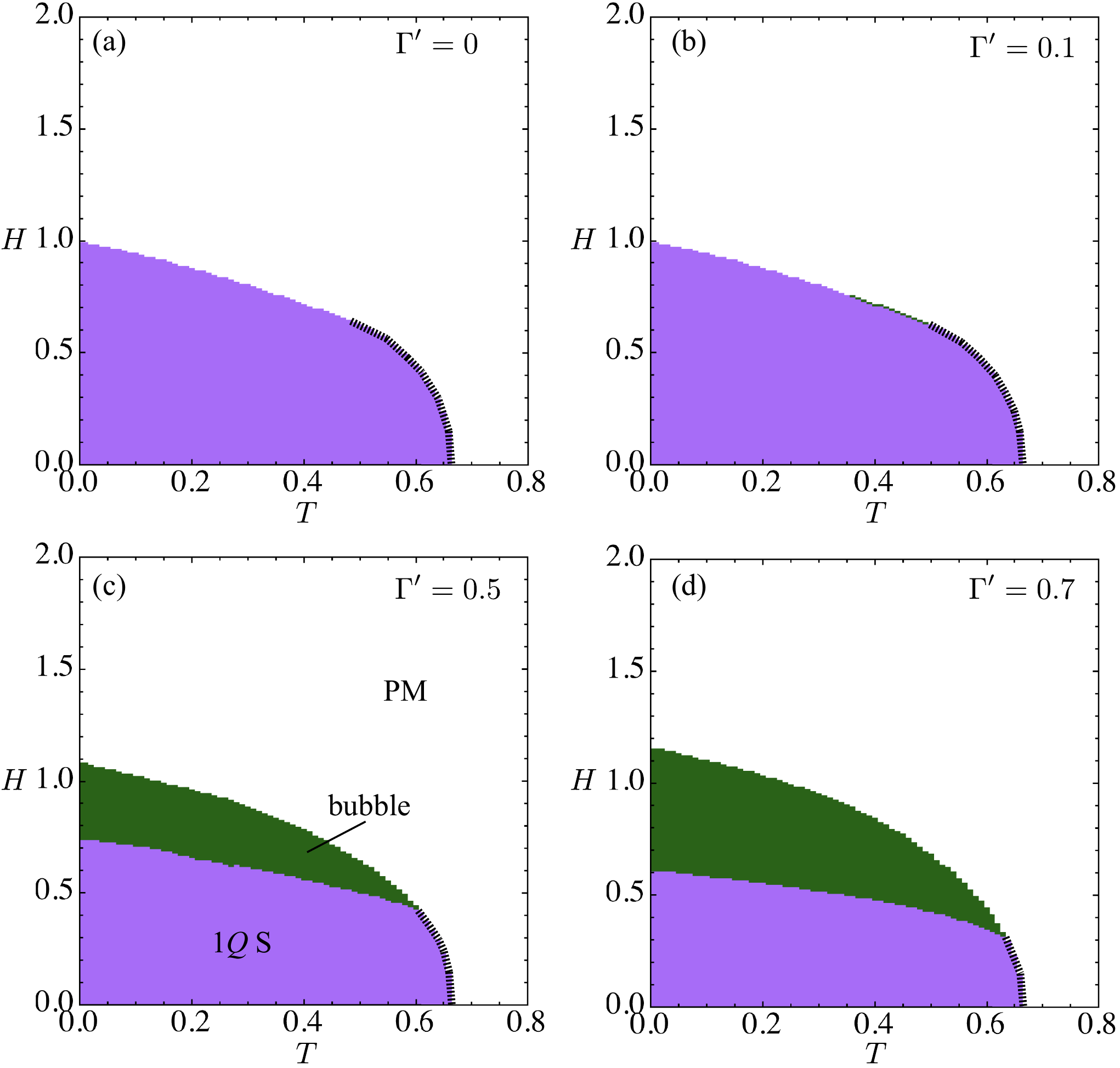} 
\caption{
\label{fig: PD_HH_2} 
$H$--$T$ phase diagrams for (a) $\Gamma'=0$, (b) $\Gamma'=0.1$, (c) $\Gamma'=0.5$, and (d) $\Gamma'=0.7$ at $K=0$ and $\kappa=0.3$. 
The dashed lines represent the second-order phase transitions.
}
\end{center}
\end{figure}

We also discuss the stability of the bubble crystal while changing $\Gamma'$ for fixed $\kappa$.
Figure~\ref{fig: PD_HH_2} shows the $H$--$T$ phase diagrams at $\kappa=0.3$ for $\Gamma'=0$ [Fig.~\ref{fig: PD_HH_2}(a)], $0.1$ [Fig.~\ref{fig: PD_HH_2}(b)], $0.5$ [Fig.~\ref{fig: PD_HH_2}(c)], and $0.7$ [Fig.~\ref{fig: PD_HH_2}(d)]; the phase diagram at larger $\kappa=0.9$ for different $\Gamma'$ is also shown in Appendix~\ref{appendix}. 
The data indicate that large (small) $\Gamma'$ tends to enhance (suppress) the region of the bubble crystal for small $\kappa$. 
In particular, no bubble crystal appears in the phase diagram at $\Gamma'=0$ in Fig.~\ref{fig: PD_HH_2}(a), which means that the high-harmonic wave-vector interaction can be a microscopic origin of the bubble crystal. 
The instability of the bubble crystal appears for $\Gamma'=0.1$ in Fig.~\ref{fig: PD_HH_2}(b), where it is stabilized in a tiny region at finite temperatures $0.63 \lesssim T \lesssim 0.75$. 
For larger $\Gamma'$ in Figs.~\ref{fig: PD_HH_2}(c) and \ref{fig: PD_HH_2}(d), the bubble crystal becomes stable even at $T=0$.

\subsection{Biquadratic interaction}
\label{sec: Biquadratic interaction}

\begin{figure*}[t!]
\begin{center}
\includegraphics[width=1.0\hsize]{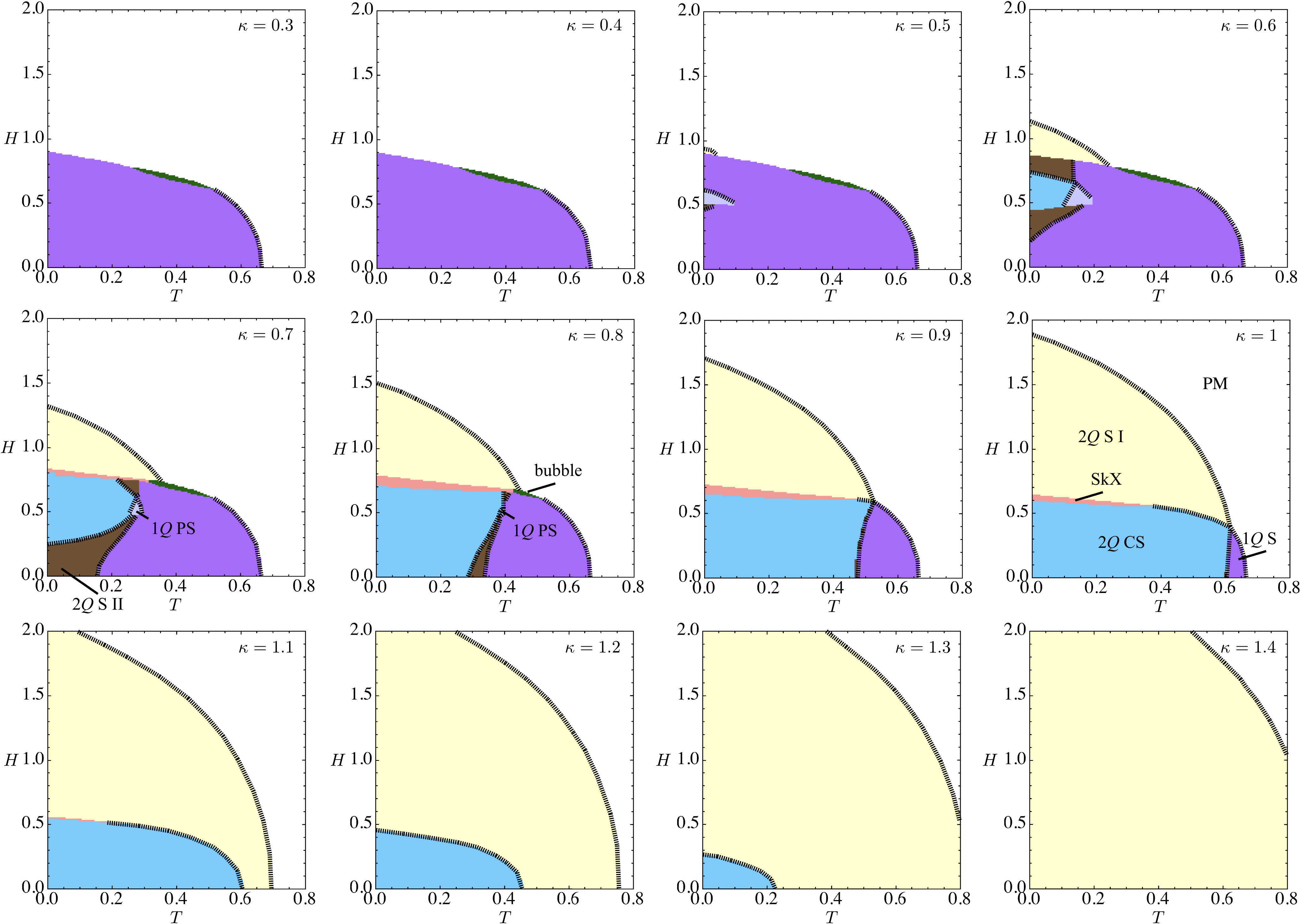} 
\caption{
\label{fig: PD_K} 
$H$--$T$ phase diagrams of the model in Eq.~(\ref{eq: Ham}) at $\Gamma'=0$ and $K=0.3$ with changing $\kappa$ for $0.3 \leq \kappa \leq 1.4$ by $\Delta \kappa =0.1$.  
The dashed lines represent the second-order phase transitions.
It is noted that the transition between the 2$Q$ CS and 2$Q$ S II states at $\kappa=0.9$ and $1$ is also second-order.
The phase diagram at $\kappa=1$ is the same as that in Ref.~\cite{hayami2023widely}. 
}
\end{center}
\end{figure*}

Next, we discuss the result under the biquadratic interaction; we set $\Gamma'=0$ in this section. 
We show a set of the $H$--$T$ phase diagrams at $K=0.3$ while changing $\kappa$ from $0.3$ to $1.4$ by $0.1$ in Fig.~\ref{fig: PD_K}. 
Since the behavior of the phases against $\kappa$ is similar to that for $\Gamma'=0.3$ and $K=0$ in Fig.~\ref{fig: PD_HH}, we discuss the similarities and differences between them by mainly focusing on the stability region of the bubble crystal. 

The $H$--$T$ phase diagram at $\kappa=1$ in Fig.~\ref{fig: PD_K} consists of five phases in addition to the PM state, which is consistent with the result in Ref.~\cite{hayami2023widely}. 
Similarly to the phase diagram under $\Gamma'=0.3$ at $\kappa=1$ in Fig.~\ref{fig: PD_HH}, the 2$Q$ CS, SkX, and 2$Q$ S I states are stabilized in the low-, intermediate-, and high-field regions from zero to finite temperatures, respectively, and the 1$Q$ S state appears in the low-field region at high temperatures next to the PM state. 
Meanwhile, there are two differences from the phase diagram for nonzero $\Gamma'$. 
One is that the 1$Q$ C and 1$Q$ PS states do not appear, since the biquadratic interaction tends to favor multiple-$Q$ states rather than the single-$Q$ states~\cite{Hayami_PhysRevB.105.174437}. 
The other is the appearance of the 2$Q$ S II state sandwiched by the 2$Q$ CS and 1$Q$ S states in the low-field region. 
The 2$Q$ S II state is characterized by a superposition of two sinusoidal waves in different spin components, i.e., $m^z_{\bm{Q}_1}, m^x_{\bm{Q}_2} \neq 0$ or $m^y_{\bm{Q}_1}, m^z_{\bm{Q}_2} \neq 0$, whose spin configuration is shown in Fig.~\ref{fig: spin}(h).

\begin{figure}[htb!]
\begin{center}
\includegraphics[width=1.0\hsize]{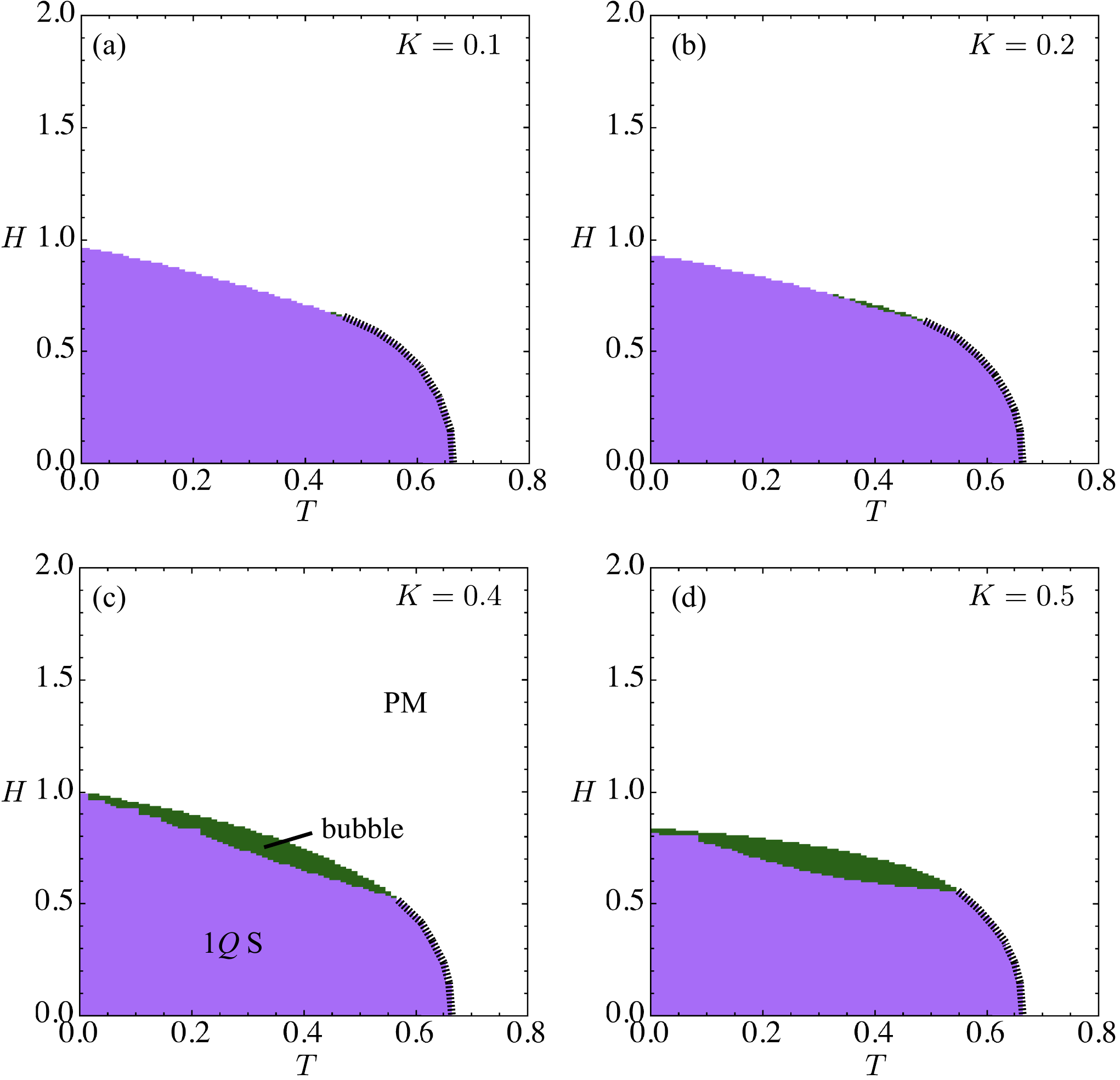} 
\caption{
\label{fig: PD_K_2} 
$H$--$T$ phase diagrams for (a) $K=0.1$, (b) $K=0.2$, (c) $K=0.4$, and (d) $K=0.5$ at $\Gamma'=0$ and $\kappa=0.3$. 
The dashed lines represent the second-order phase transitions.
}
\end{center}
\end{figure}

When we set $\kappa<1$, the bubble crystal appears in the phase diagrams for $\kappa \lesssim 0.8$, as shown in Fig.~\ref{fig: PD_K}, whose stability tendency is similar to that in the case of $\Gamma' \neq 0$ in Sec.~\ref{sec: Higher-harmonic wave-vector interaction}; it is stabilized in the intermediate-field and finite-temperature region next to the SkX. 
Thus, the finite-temperature phase transition is also expected in the mechanism based on the biquadratic interaction.

The results indicate that the interplay between $K$ and $\kappa$ can become another microscopic mechanism of the bubble crystal.  
In fact, the region of the bubble crystal becomes narrower as $K$ decreases, as shown in the phase diagrams at $K=0.1$ in Fig.~\ref{fig: PD_K_2}(a) and at $K=0.2$ in Fig.~\ref{fig: PD_K_2}(b). 
On the other hand, its region is extended for larger $K$ as shown in the cases of $K=0.4$ in Fig.~\ref{fig: PD_K_2}(c) and $K=0.5$ in Fig.~\ref{fig: PD_K_2}(d); the bubble crystal remains stable at zero temperature for $K=0.5$. 

For $\kappa>1$, the 1$Q$ S and 2$Q$ S II states are replaced by the 2$Q$ S I state.  
In addition, the regions for the SkX and 2$Q$ CS states become narrower as $\kappa$ increases, as shown in the phase diagrams in Fig.~\ref{fig: PD_K}. 
In the end, all the regions are occupied by the 2$Q$ S I state at $\kappa = 1.4$, which is different from the situation in Fig.~\ref{fig: PD_HH}, where the 1$Q$ C state appears besides the 2$Q$ S I state.

\section{Summary}
\label{sec: Summary}

To summarize, we have investigated the stabilization mechanisms of the bubble crystal in tetragonal magnets. 
We constructed the magnetic field--temperature phase diagrams while changing the easy-axis magnetic anisotropic two-spin interaction in a systematic way by the efficient steepest descent method. 
We found that there are two key ingredients in addition to the easy-axis two-spin magnetic anisotropy to realize the bubble crystal: the high-harmonic wave-vector interaction and the biquadratic interaction. 
We showed that the bubble crystal phase appears in the high-temperature region of the SkX phase for a moderate easy-axis anisotropic two-spin interaction; they are transformed into each other by changing the temperature. 
In addition, we showed that the bubble crystal remains stable in the zero-temperature limit for the large high-harmonic wave-vector interaction or biquadratic interaction in addition to the large easy-axis anisotropic two-spin interaction. 
Meanwhile, the bubble crystal is destabilized when the magnetic anisotropic two-spin interaction tends to be easy-plane. 

The present result indicates that the instability toward the bubble crystal can occur in the high-temperature region to the SkX phase. 
Thus, there is a chance of inducing the bubble crystal in the skyrmion-hosting materials, such as GdRu$_2$Si$_2$~\cite{khanh2020nanometric, Yasui2020imaging, khanh2022zoology, Matsuyama_PhysRevB.107.104421} and EuAl$_4$~\cite{takagi2022square, Gen_PhysRevB.107.L020410}, if the easy-axis two-spin magnetic anisotropy is strengthened by applying an external pressure and chemical substitution. 
Furthermore, our model can be also applied to materials with strong easy-axis two-spin magnetic anisotropy so that the SkX is not stabilized. 
CeAuSb$_2$ is one of the candidates, where the magnetic field--temperature phase diagrams in Figs.~\ref{fig: PD_HH_2}(c) and \ref{fig: PD_HH_2}(d) resemble that observed in experiments~\cite{Marcus_PhysRevLett.120.097201,Park_PhysRevB.98.024426,Seo_PhysRevX.10.011035,seo2021spin}.

\begin{acknowledgments}
This research was supported by JSPS KAKENHI Grants Numbers JP21H01037, JP22H04468, JP22H00101, JP22H01183, JP22K03509, JP23K03288, JP23H04869, and by JST PRESTO (JPMJPR20L8). 
Parts of the numerical calculations were performed in the supercomputing systems in ISSP, the University of Tokyo.
\end{acknowledgments}

\appendix
\section{Phase diagram for different $\Gamma'$}
\label{appendix}

\begin{figure}[tb!]
\begin{center}
\includegraphics[width=1.0\hsize]{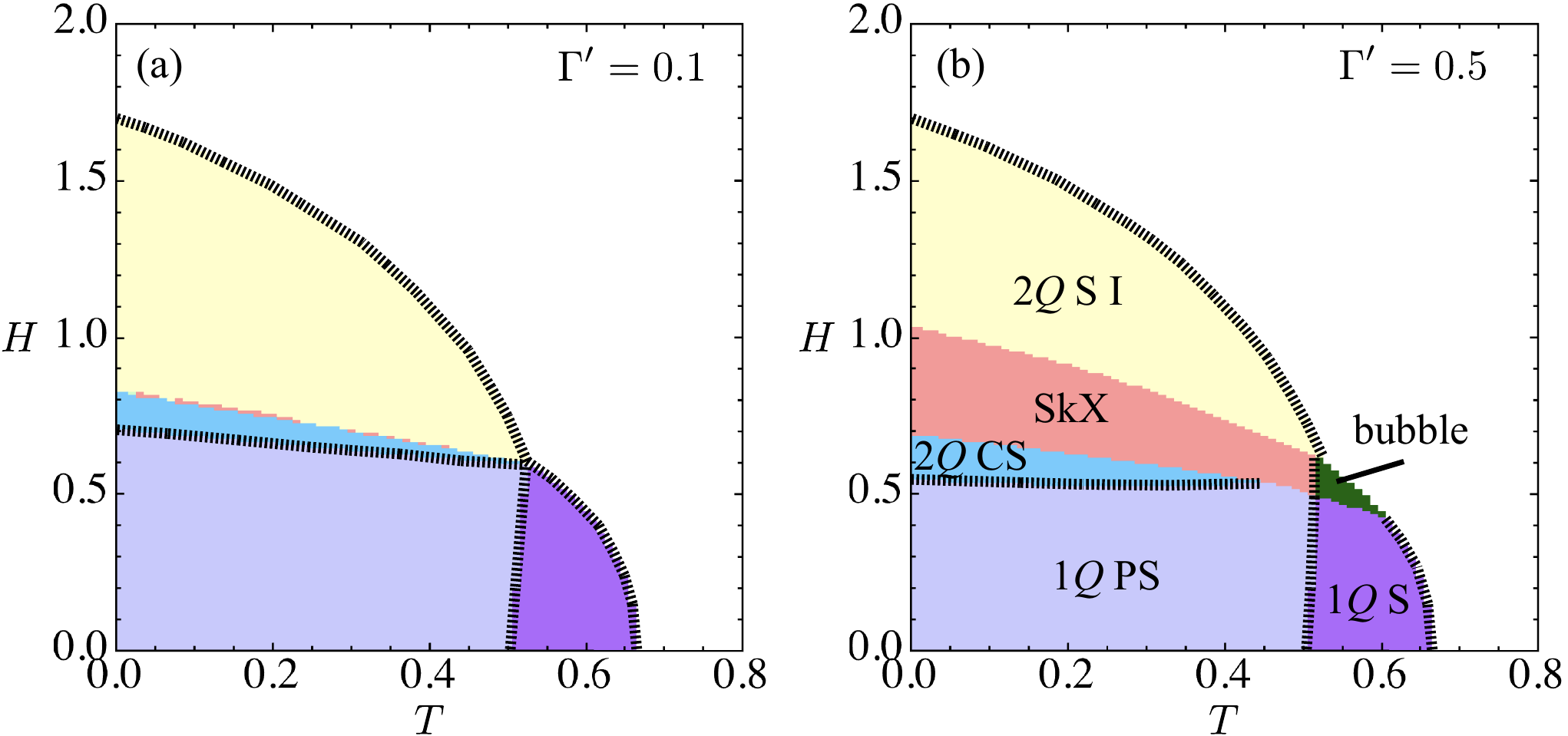} 
\caption{
\label{fig: PD_HH_3} 
$H$--$T$ phase diagrams of the model in Eq.~(\ref{eq: Ham}) at $K=0$ and $\kappa=0.9$ for (a) $\Gamma'=0.1$ and (b) $\Gamma'=0.5$. 
The dashed lines represent the second-order phase transitions.
It is noted that the phase transition between the 2$Q$ CS and 2$Q$ S I states in (a) is also second-order except for the region where the SkX appears between them. 
}
\end{center}
\end{figure}

Figure~\ref{fig: PD_HH_3} shows the $H$--$T$ phase diagram at $K=0$ and $\kappa=0.9$ for different $\Gamma'$; the result for $\Gamma'=0.1$ ($\Gamma'=0.5$) is shown in Fig.~\ref{fig: PD_HH_3}(a) [Fig.~\ref{fig: PD_HH_3}(b)]. 
By referring to the result at $K=0$, $\kappa=0.9$, and $\Gamma'=0.3$ in Fig.~\ref{fig: PD_HH}, one can find that the region of the bubble crystal is shrunk (extended) with decreasing (increasing) $\Gamma'$, which is a similar tendency for small $\kappa$ in Fig.~\ref{fig: PD_HH_2} in Sec.~\ref{sec: Case of other Gamma}. 
In addition, one notices that the phase boundary between the SkX and the bubble crystal is almost unchanged by $\Gamma'$, which indicates that the energy gain by $\Gamma'$ is comparable between them.  
These results suggest that the bubble crystal can be robustly stabilized in the high-temperature region next to the SkX even for less anisotropy $\kappa \simeq 1$ when the contribution from the high-harmonic wave-vector interaction becomes large.

\bibliographystyle{apsrev}
\bibliography{ref}

\begin{thebibliography}{68}
\bibitem[{com()}]{comment_bubble_PM}
\bibinfo{note}{At the phase boundary on the high-temperature side of the bubble
  crystal, the jumps of $M^z$ and $m^z_{\bm{Q}_{1,2,3,4}}$ is smaller and the
  order of the transition is difficult to determine.}
\end{thebibliography}
\end{document}